\documentclass[AMA,STIX1COL]{WileyNJD-v2}
\usepackage{moreverb}

\newcommand\BibTeX{{\rmfamily B\kern-.05em \textsc{i\kern-.025em b}\kern-.08em
T\kern-.1667em\lower.7ex\hbox{E}\kern-.125emX}}

\articletype{Journal}%
    
\received{<day> <Month>, <year>}
\revised{<day> <Month>, <year>}
\accepted{<day> <Month>, <year>}

\usepackage{amsmath,amssymb,amsfonts}
\usepackage{graphicx}
\usepackage{textcomp}
\usepackage[utf8]{inputenc}
\usepackage{xcolor}
\usepackage{listings}
\usepackage{caption}
\usepackage{todonotes}
\usepackage{subcaption}
\usepackage[shortlabels]{enumitem}
\usepackage{url}
\usepackage{mdframed}
\usepackage{ulem,xpatch}
\usepackage{lineno,hyperref}

\lstdefinestyle{top}{
  float=tp,
  floatplacement=tbp,
}

\lstdefinelanguage{docker}{
	keywords={FROM, RUN, COPY, ADD, ENTRYPOINT, CMD,  ENV, ARG, WORKDIR, EXPOSE, LABEL, USER, VOLUME, STOPSIGNAL, ONBUILD, MAINTAINER},
	keywordstyle=\color{blue}\bfseries,
	identifierstyle=\color{black},
	sensitive=false,
	comment=[l]{\#},
	commentstyle=\color{purple}\ttfamily,
	stringstyle=\color{red}\ttfamily,
	morestring=[b]',
	morestring=[b]"
}

\lstdefinelanguage{docker-compose}{
	keywords={image, environment, ports, container_name, ports, volumes, links},
	keywordstyle=\color{blue}\bfseries,
	identifierstyle=\color{black},
	sensitive=false,
	comment=[l]{\#},
	commentstyle=\color{purple}\ttfamily,
	stringstyle=\color{red}\ttfamily,
	morestring=[b]',
	morestring=[b]"
}
\lstdefinelanguage{docker-compose-2}{
	keywords={version, volumes, services},
	keywordstyle=\color{blue}\bfseries,
	keywords=[2]{image, environment, ports, container_name, ports, links, build, depends_on},
	keywordstyle=[2]\color{olive}\bfseries,
	identifierstyle=\color{black},
	sensitive=false,
	comment=[l]{\#},
	commentstyle=\color{purple}\ttfamily,
	stringstyle=\color{red}\ttfamily,
	morestring=[b]',
	morestring=[b]"
}

\newcommand\red[1]{\textcolor{black}{#1}}

\begin{document}

\title{Microservice Architecture Practices and Experience: a Focused Look on Docker Configuration Files}

\author[1]{Luciano Baresi}

\author[1]{Giovanni Quattrocchi}
\author[2]{Damian Andrew Tamburri}

\address[1]{Dipartimento di Elettronica, Informazione e Bioingegneria\\Politecnico di MIlano, Milan, Italy}

\address[2]{
Jheronimus Academy of Data Science, Tilburg University, The Netherlands
}

\authormark{Baresi, Quattrocchi, Tamburri}

\corres{*Giovanni Quattrocchi, Corresponding address. \email{giovanni.quattrocchi@polimi.it}}

\abstract[Abstract]{
Cloud applications are more and more microservice-oriented, but a concrete charting of the microservices architecture landscape---namely, the space of technical options available for microservice software architects in their decision-making---is still very much lacking, thereby limiting the ability of software architects to properly evaluate their architectural decisions with sound experiential devices and/or practical design principles. 
On the one hand, Microservices are fine-grained, loosely coupled services that communicate through lightweight protocols. 
On the other hand, each microservice can use a different software stack, be deployed and scaled independently or even executed in different containers, which provide isolation and a wide-range of configuration options but also offer unforeseeable architectural interactions and under-explored architecture smells, with such experience captured mainly in software repositories where such solutions are cycled. 

This paper adopts a mining software repositories (MSR) approach to capture the practice within the microservice architecture landscape,  by eliciting and analysing Docker configuration files, being Docker the leading technical device to design for, and implement modern microservices. Our analysis of Docker-based microservices gives an interesting summary of the current state of microservices practice and experience. Conversely, observing that all our datapoints have their own shape and characteristics, we conclude that further comparative assessment with industrial systems is needed to better address the recurring positive principles and patterns around microservices. 
}


\maketitle

\section{Introduction} 
\label{sec:intro}

In little more than 5 years \textit{microservices}~\cite{dragoni2017microservices} were proposed and aggressively expanded as a novel architectural style to conceive and operate complex cloud systems (e.g., Netflix~\cite{Netflix}). While some say microservices ``simply" reflect SOA (Service Oriented Architecture \cite{mackenzie2006reference}) done right, both practice and experience show there are several technical differences, e.g., services originally had the goal of increasing reuse and cooperation, while \textit{micro}services foster independence and scalability. Furthermore, microservice applications are conceived as multitudes of loosely-coupled small---or micro, i.e., \emph{single functional units}---services that manage their own data and communicate through simple REST interfaces or even lightweight publish-and-subscribe buses~\cite{lewis2014microservices,nadareishvili2016microservice}. 

On the one hand, each microservice is (a) typically based on a different software stack, (b)  developed by an independent team, (c) run independently of the others in an isolated execution environment, and (d) deployed, managed, and scaled freely, thereby minimising overprovisioning and infrastructure costs as well as risks.

On the other hand, containers~\cite{kang2016container} are the key lightweight virtualization technology that operates microservices. A container wraps a process and provides an isolated execution environment that can be managed independently of the other processes, with Docker being the de-facto practice device for microservice operations\footnote{\url{https://www.docker.com}}.Docker uses declarative configuration files to state how to create a container image and to define how different containers can be deployed and connected to one another.

With the lack of a map to chart the aforementioned architecture practices and experiences in their respective software architecture landscape \cite{cervantes2016designing}, the process of architecture decision-making for microservices computing---i.e., making and changing architecture decisions as part of architectural design \cite{bassSoftwareArchInPractice}---becomes even more complex and risky, reflecting little or no decision support, especially so for novice architects. 


To address this gap, we perform a Mining Software Repositories (MSR) \cite{PoncinSB11} study focusing on Docker, with the aim of gathering and representing design principles and available practices drawn from the practical experience of docker-focused microservice architects out there. Specifically, our analysis features a mining tool of our own design that searched and downloaded  microservice-related repositories in GitHub that contain Docker configuration files of two types: \textit{Dockerfiles} and \textit{docker-compose} files. The former files are used to define how single components/microservices should be installed, configured and run, while the latter are used to specify how different components/microservices are connected one another. Even though most of real-world microservice-based applications are closed-source and not available for our analysis, we carefully selected, through a semi-automated pipeline, 29 open-source projects from GitHub. We advocate that these applications can help researchers and practitioners understand what are the most common practices employed for the development microservices. 

 	

The analysis of Docker configuration files reveals that while some well-known practices and principles are quite established in the microservice landscape---e.g., the tendency of having small, independent components for under 3 business functions---others are not, quite the contrary in fact; for example, our analysis of the practice shows that the usage of multiple programming languages and software stacks at the same time is negligible. Besides showcasing this and similar experiential findings, the paper discusses open issues reported while analyzing collected data which require further investigations.

The rest of the paper is organized as follows. Section~\ref{sec:background} introduces Docker and its configuration files. Section~\ref{sec:rq} summarizes the main characteristics of microservice architectures and introduces our research questions. Section~\ref{sec:methods} describes how we searched GitHub repositories and what we retrieved. Section~\ref{sec:results} presents obtained results and Section~\ref{sec:discussion} discusses them. Section~\ref{sec:related} compares our findings against related work, and Section~\ref{sec:conclusions} concludes the paper. 

\section{Microservices Computing: The Docker Approach} 
\label{sec:background}
\begin{figure}
     \centering
     \begin{subfigure}[b]{0.48\columnwidth}
         \centering
         \includegraphics[width=\textwidth]{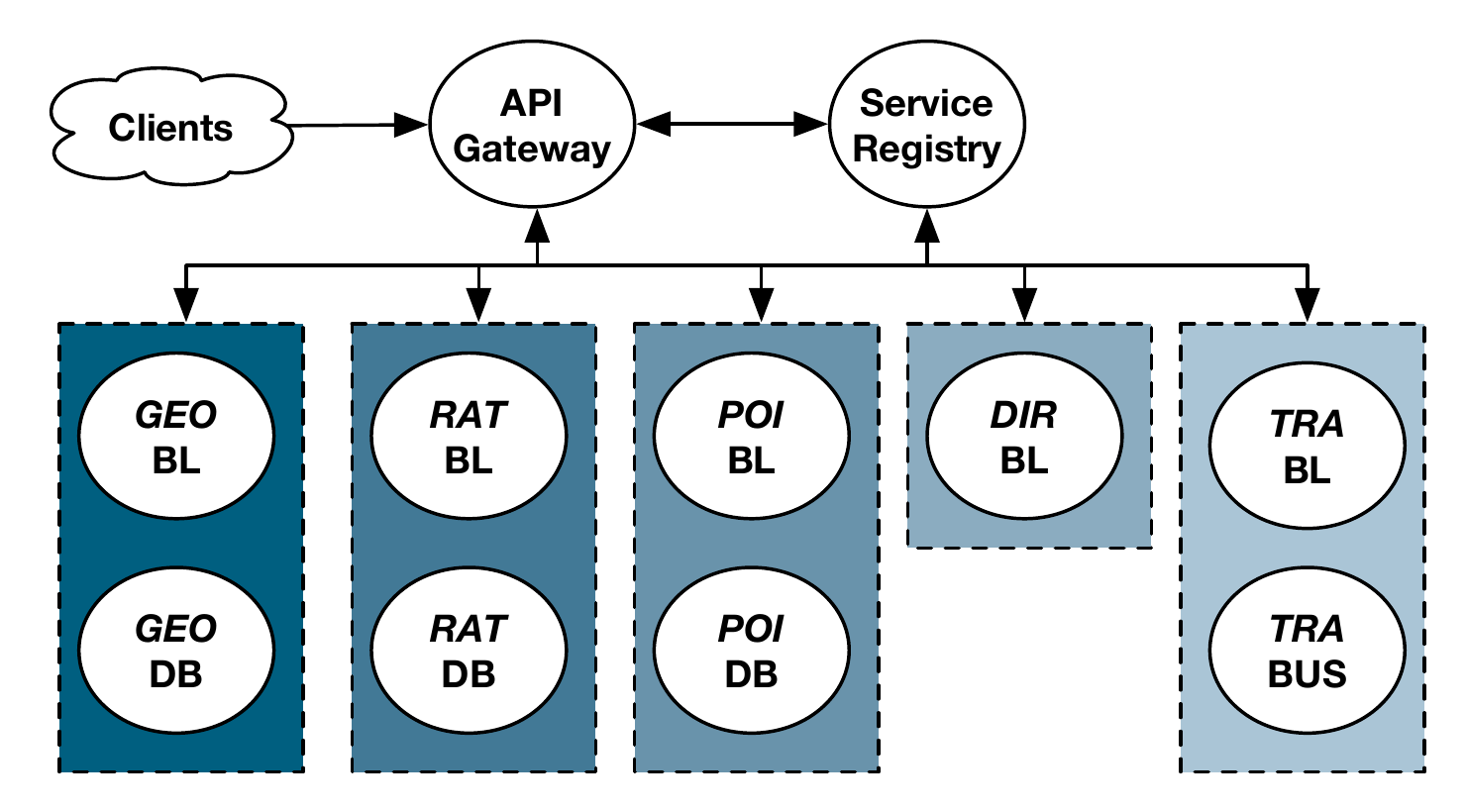}
         \caption{Architecture}
         \label{fig:example:arch}
     \end{subfigure}
     \begin{subfigure}[b]{0.48\columnwidth}
         \centering
         \includegraphics[width=\textwidth]{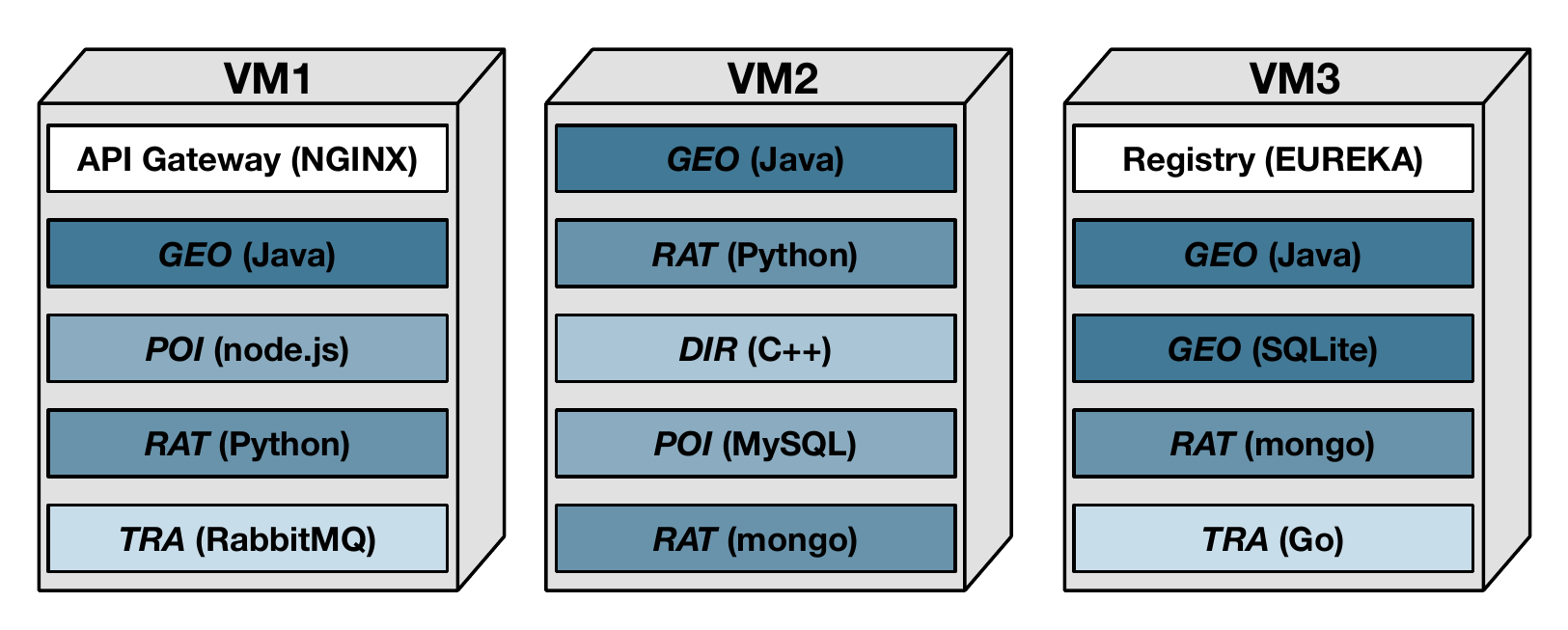}
         \caption{Deployment}
         \label{fig:example:dep}
     \end{subfigure}
     \hfill
        \caption{\textit{MAP} application.}
         \label{fig:example}
\end{figure}

Microservices are usually deployed using containers to get isolation and independent management. A single microservice is usually deployed using multiple containers since it often exploits external components, like a load balancer or an external DBMS, and then each component is run in a separate container. The same microservice can also be instantiated multiple times to serve multiple clients. Different instances can be distributed onto a cluster of physical/virtual machines that are shared with other microservices and related components. 

To exemplify the aforementioned characteristics and the flexibility provided by microservices and containers, Figure~\ref{fig:example:arch} shows the architecture of \textit{MAP},  an exemplar application composed of five different microservices (dashed colored boxes): \textit{GEO}, \textit{POI}, \textit{RAT}, \textit{DIR}, and \textit{TRA}.  Each microservice comprises a business logic (BL) component and, in if needed, a DBMS (DB) or a bus infrastructure (BUS).
Figure~\ref{fig:example:dep} proposes a possible container-based deployment of 
\textit{MAP} onto a cluster of three VMs: each black-bordered rectangle is a container and each web server, DBMS, and bus infrastructure is deployed using a separated container.

\textit{GEO} is a microservice written in Java, exploits an sqlite database to store the geographical elements of maps, and exposes a REST API through the Spring framework. Since \textit{GEO} is the most widely used service in the system, it is replicated on all the three VMs. \textit{POI} is dedicated to retrieving point of interests, is written Javascript (node.js), and exploits a MySQL database.  \textit{RAT} is used to manage and rate the interactions between customers and POIs, is written in Python, and exploits a gunicorn web server and a distributed  mongo database. \textit{RAT} is deployed using two replicas for the web server and two containers for the distributed database.  \textit{DIR} is dedicated to retrieving directions given source and destination addresses. It uses a C++ optimized algorithm to compute the best path and is stateless. Finally, \textit{TRA} is dedicated to traffic updates, is written in Go, and uses \textit{RabbitMQ} to send notifications to clients.  In addition to the microservices, we also deployed an instance of an NGINX API Gateway\footnote{\url{https://www.nginx.com}} and of a Netflix Eureka\footnote{\url{https://github.com/Netflix/eureka}} service registry to ease client interactions and facilitate service discovery.

\subsection{Docker configuration files}

\begin{figure}[t]
\begin{minipage}[b]{.5\textwidth}
\begin{lstlisting}[language=docker,caption={\textit{Dockerfile} of $RAT$.},breaklines=false,label={lst:dockerfile}, numbers=left, xleftmargin=0.8cm, belowskip=1em]
FROM ubuntu:16.04
WORKDIR /app
COPY ./src /app
RUN apt-get update -y && \
apt-get install gunicorn && \
apt-get install -y python-pip && \ 
apt-get install -y python-dev && \
pip install -r requirements.txt
CMD [ "python", "RAT.py" ]
\end{lstlisting}
\end{minipage}\hfill
\begin{minipage}[b]{.5\textwidth}
\begin{lstlisting}[language=docker-compose-2,caption={ docker-compose file of \textit{MAP}.}, breaklines=false,label={lst:compose}, numbers=left, xleftmargin=0.8cm,belowskip=1em]
services:
  NGINX:
    image: nginx:1.13-alpine
    depends_on:
      - EUREKA
  EUREKA:
   image: netflixoss/eureka
  RAT:
    build: ./microservices/RAT
    depends_on:
      - MONGO_RAT
      - EUREKA
  MONGO_RAT:
    image: mongo
    environment:
      - MONGO_INIT_DATABASE=ratdb
  GEO:
    build: ./microservices/GEO
  ...
\end{lstlisting}
\end{minipage}
\end{figure}

As said, the empirical study presented in this paper is based on two types of Docker files: \textit{Dockerfiles} and \textit{docker-compose} files. Each Dockerfile file contains a set of instructions that install the required software to build an image. These files are also used to transfer files from local drives to container images, expose container ports, and define the startup command that lunches the embedded process. Listing~\ref{lst:dockerfile} shows a hypothetical Dockerfile for $RAT$. The \textit{FROM} statement of line $1$ declares the base image for the container. In this case, \textit{RAT} is based on \textit{ubuntu} version $16.04$. This means that on top of the shared kernel, the aforementioned ubuntu distribution will be installed into the container image. Note that the image specified through \textit{FROM} must be present in a registry of container images, where the default is Docker Hub\footnote{\url{https://hub.docker.com}}. , the public Docker registry.

The \textit{WORKDIR} statement of line $2$ defines the folder of the target container image onto which subsequent instructions will be executed. If the folder does not exist it is created during the building process. The \textit{COPY} statement at line $3$ loads a local disk folder onto the container image.  In this case, folder \textit{src} is loaded onto \textit{app}. The \textit{RUN} statement (lines $4-8$) defines the bash commands that are executed during the building process. These commands are mostly used to install the required dependencies to build the container image. The example statement first updates package manager \verb#apt-get#, and then installs \textit{gunicorn}, python and \textit{pip}\footnote{\url{https://packaging.python.org/guides/tool-recommendations}}. These dependencies are explicitly defined in the \textit{Dockerfile}, meaning that the needed software is stated explicitly. 

External python modules are installed through the command at line $8$. In this case, the dependencies are not explicitly defined in the Dockerfile, but they are specified in an external file, \verb#requirements.txt#, which is loaded into the container image with the \textit{COPY} statement at line $3$.  Finally, the \textit{CMD} statement of line $9$ defines the default command to execute to run an instance of the created image. The example launches a python file that initializes \textit{RAT}.

\textit{docker-compose} files are used to define the sets of containers to be deployed in a declarative way using YAML. Listing~\ref{lst:compose} shows an example file for deploying \textit{MAP}. The $services$ statement contains an associative array where keys are the names (decided by the user) of the services to be deployed and values are the configuration parameters of each service. A service in \textit{docker-compose}, or a \textit{compose} service, does not directly refer to a microservice since it is only bound to one container image, while a microservice can exploit multiple ones. Listing~\ref{lst:compose}  only shows five services out of the eleven introduced above.

A \textit{docker-compose} service can contain different attributes. Attribute \textit{image} is used when the container image is already published in the used registry. For example, services \textit{NGINX} (line $3$), \textit{EUREKA} (line $7$) and \textit{MONGO\_RAT} (line $14$) use pre-built images available in Docker Hub. In contrast, services \textit{RAT} and \textit{GEO} use attribute \textit{build} (at line $9$ and $18$, respectively) to state the path of the Dockerfile to build to create the proper images.

Attribute \textit{environment} is used to define environment variables within the container. For example, a container that instantiates service  \textit{MONGO\_RAT} embeds an environment variable \textit{MONGO\_INIT\_DATABASE} (line $16$) with value \textit{ratdb} (the name of the used mongo database). 
 
Finally, attribute \textit{depends\_on} is used to define the relationships among services. For example, to ease service discovery, we can assume that all services, but the DBs, depend on the Eureka service registry: Listing~\ref{lst:compose} defines the dependency at line $4$ for \textit{NGINX} and at line $12$ for \textit{RAT}. In addition \textit{RAT} depends on its dedicated mongo database (service \textit{MONGO\_RAT}) as shown at line $11$. 

\section{Research Problem \& Questions}
\label{sec:rq}
Practitioners rely on a set of well-known background principles~\cite{lewis2014microservices}---recapped below---when designing a new microservice architecture or migrating a legacy system towards a microservice-oriented design. Such background helped us formulate a concrete research problem, subsequent questions, and operationalize them.

\subsection{Background and Research Problem}
First, even if nobody has defined the maximum size of a microservice, practitioners converge towards a principled definition of size, agreeing that the component should be \textit{small enough} to be developed and managed with agility. A microservice should be built around a single functionality. For example, in a mapping application, a microservice can be dedicated to retrieving geographical portions on maps, another to searching for points of interest (POI), and another to finding directions. A reduced size is also key for automation. Continuous integration and delivery~\cite{o2017continuous} can benefit from smaller projects since they are simpler and faster to manage.

Subsequently, Microservices are usually built and managed \textit{independently} one another. The team in charge of a microservice should be small and include enough, heterogeneous skills to cover the whole life-cycle. Amazon reported that the biggest team to handle a single service should be around a dozen people (Two Pizza Team rule~\cite{awspizza}).

At the same time, each microservice can be built using a different technology and programming language.  Instead of single solutions that must fit all needs, these \textit{polyglotism} and flexibility help address particular tasks and needs. This also means that each team can deliver updates as if its microservice were an \textit{independent product}. 

What is more, Microservices can also exploit auxiliary frameworks and DBMSs. Note that each microservice should refer to an \textit{independent data store}. While this characteristic is coherent with the idea of loosely-coupled components, some data can be duplicated, most transactions become distributed, and the overhead of retrieving proper data could significantly impact the performance of the system. Dedicated \textit{patterns}, such as the Saga pattern~\cite{vstefanko2019saga}, can be used to enforce consistency among different local transactions.

Also, Microservices should make use of \textit{lightweight communication}. While SOAP services rely on SOAP itself and Enterprise Service Busses (ESB) embed complex communication capabilities, microservices only exploit HTTP and offer simple REST interfaces. The interactions are then mostly asynchronous and based on messaging over a dumb, \textit{message bus} (similar to Unix pipes) based on MQTT\footnote{\url{https://www.iso.org/standard/69466.html}} (e.g., RabbitMQ\footnote{\url{https://www.rabbitmq.com}}).
The dimension of microservices helps keep internal, hidden dependencies under control; inter-service communication does not exploit complex orchestrators (e.g., a BPEL engine~\cite{fu2004analysis}). Microservices are choreographied and complexity is managed by the services' end points that produce and consume messages. Chatty communications, like the usual interactions through method calls, are substituted by more coarser-grained ones, and this is why the end points become the smart part of the system. 

Finally, while microservices split the business logic into small units, clients could need information that exceed the boundary of a single call. To avoid shifting complexity to clients, practitioners rely on \textit{API gateway}s~\cite{taibi2018architectural}. An API gateway is a dedicated component that exposes coarse grained endpoints to clients and translate the request into multiple calls to the proper microservices. An API gateway can exploit a \textit{Service registry}, which stores information about available microservices (e.g., their IP addresses), to perform the calls properly. Service registries could also be used by clients directly to easily locate services, be informed about their health status, and orchestrate the calls. Part of the complexity of using different technologies is removed by containers. They can easily be reused and provide a single interface for managing running components.

In line with the background principles above, the research problem we address is the following:

\begin{mdframed}
\textbf{Problem Statement.} Several established principles exist in the state of practice around Microsevice design and computing. Conversely, little is known on whether such principles actually reflect the technical design decisions made by microservice architects in their practice. 
\end{mdframed}

\subsection{Research questions}

These principles and characteristics led us to identify the following research questions: 

\noindent\textbf{RQ1 - Sizes.} How big are microservice projects? How many microservices are there in a project? How big are the teams that work on microservice projects?

\noindent\textbf{RQ2 - Polyglotism.} Which are the most used programming language? How spread is the usage of multiple languages in the same project? How are they combined?

\noindent\textbf{RQ3 - Application servers and DBMSs.} What are the most widely used application servers and DBMSs in microservice projects? Are independent data stores reality?

\noindent\textbf{RQ4 - Topologies and Data Independence.} How dependent are microservices on one another? Are independent data stores reality?

\noindent\textbf{RQ5 - Communication and management.}  Do microservices communicate through bus infrastructures? How frequent is the usage of API gateways and service registries? Are microservice architectures monitored?
 
\noindent\textbf{RQ6 - Reused container images.} What are the most used container images in microservice projects?

\section{Pipeline and Dataset Descriptive Statistics}
\label{sec:methods}

\begin{figure}[t]
    \centering
	\includegraphics[width=0.65\columnwidth]{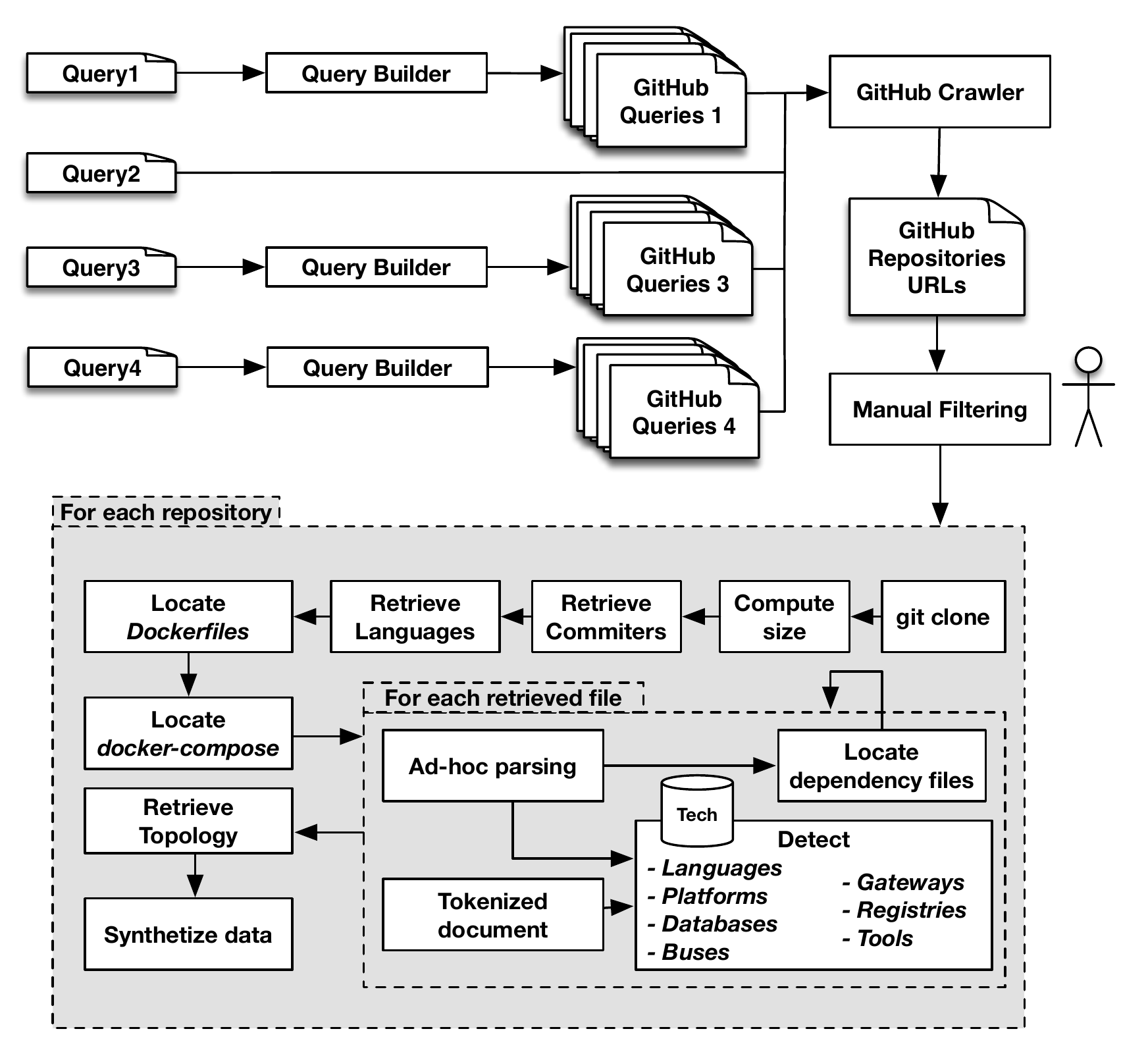}
	\caption{Pipeline used to carry out the assessment. }
	\label{fig:pipeline}
\end{figure}

Figure~\ref{fig:pipeline} shows the pipeline we created to retrieve the dataset and carry out the analyses. As first step, we created four queries to retrieve the repositories that are related to microservices and contain Docker configuration files from GitHub\footnote{\url{https://docs.github.com/en/rest}} (using API version 3). The first query retrieved repositories with keywords \textit{microservice} and \textit{docker} in their metadata and provided \red{$6,594$} results. The second query retrieved repositories with keywords \textit{microservice} and \textit{container} in their metadata and provided \red{$952$} results. The third query retrieved repositories with a Dockerfile that contains keyword \textit{microservice} and provided \red{$3,602$} results. The fourth query retrieved repositories with docker-compose files that contain keyword \textit{microservice} and produced \red{$2,191$} results.

GitHub enforces the limit that each query can return a maximum of $1,000$ results. To bypass this limitation, we created a \textit{Query Builder} that, given a query $Q$, generates multiple sub-queries that return less than $1,000$ results each and aggregates the sub-results to obtain the complete outcome of $Q$. The \textit{Query Builder} adds to the original queries appropriate selectors that further constrain the scope of the search. For the first two queries, it uses GitHub parameter \textit{created} to generate multiple sub-queries that inspect different suitable date intervals. For the last two queries, it uses parameter \textit{size} to constrain the search to files with a size greater or smaller than a given threshold. In total, the Query Builder generated $10$ queries that were executed by \textit{GitHub Crawler}, a component based on  \texttt{gitpython} that we built\footnote{\url{https://github.com/deib-polimi/github-crawler}} to facilitate querying GitHub repositories. The output is the list of the URLs of all repositories that match the query. 

The four queries retrieved \red{$13,339$} unfiltered repositories. We then removed the duplicates, the repositories with less than $10$ stars, and the ones with a last commit before \red{October 1, 2021}, to keep ``quality'' and active repositories~\cite{7816479}. The resulting dataset contained \red{$184$} repositories \red{($1.38\%$)}. 
 
We then started analyzing these repositories, but we obtained unrealistic results (see Section~\ref{sec:threats}). Therefore, we inspected the \red{$184$} repositories by hand. We manually categorized each repository using one of the following labels: i) $M$ \red{($7.5\%$)}, proper and complex microservice architecture; ii) $D$ \red{($27.4\%$)}, proper and complex microservice architecture uploaded for demo purposes or as reference application; iii) $S$ \red{($23.3\%$)}, single microservice or dedicated tool to be deployed along with an application (e.g., the repository of an API gateway); iv) $T$ \red{($18.5\%$)}, an external microservice tool (e.g., a management platform); v) $B$ \red{($8.9\%$)},  a repository connected to a simple tutorial, guide, book, or workshop about microservices; and iv) $O$ \red{($14.4\%$)}, a repository with poor or no relation with microservices. 
 
While the queries were able to select microservice-related repositories, more than half of them were unrealistic instances of microservice architectures, tools, or small examples (labels $S$, $T$, $B$). We then further restricted the dataset by only selecting $M$- and $D$-labelled repositories and obtained a final dataset of \red{$51$} results.

For each repository in the dataset, we cloned the code base locally and analyzed it with a dedicate tool\footnote{\url{https://github.com/deib-polimi/microservices-analysis}} we created. We first computed the project size without considering unnecessary data (e.g., folder \texttt{.git}).
Then, we retrieved the number of contributors to the repository using command \textit{git log}. We used \textit{github-linguist}\footnote{https://github.com/github/linguist}, a library maintained by GitHub itself, to detect the programming languages used in the repository, along with their percentage of usage in the source code. We also filtered out obtained results to exclude all the languages with an utilization lower than 10\% and all markup languages (e.g., HTML or CSS). 

We then retrieved  all the Dockerfiles included in each repository and the docker-compose file (if present). 
For each file retrieved, we applied a tokenization step to remove all punctuation symbols, digits, and special characters. In addition, we parsed the Dockerfiles and docker-compose files to extract the key parts of each file. We extracted the content of \textit{FROM}, \textit{RUN}, \textit{COMMAND} statements of Dockerfiles and of \textit{services}, \textit{image}, and \textit{build} statements of docker-compose files. In the tokenized document and in the output of the parsing step, we searched for programming languages, platform components, such as web-servers, databases, bus infrastructures, gateways, service registries, and other related tools. To do that we matched each retrieved datum with a set of well-known technologies retrieved manually from the Internet\footnote{A complete list of used technologies can be found here: \url{https://github.com/deib-polimi/microservices-analysis/tree/master/consts.}}. 

In addition to Docker configuration files, we located all the possible auxiliary files that could contain significant references to dependencies. More specifically, we located i) the files directly mentioned in one of the retrieved Docker configuration files (e.g., \verb#requirements.txt# at line $8$ of Listing \ref{lst:dockerfile}), and ii) for all detected languages/technologies, well-known additional dependency files such as Maven or Gradle files. The same analysis step was then carried out for these additional files.

Finally, we used docker-compose files to retrieve the topology graph of the project by constructing the graph using attribute \textit{depends\_on} of each service. This way, we could understand how each microservice is connected to the others. For each repository, we then synthesized the data by aggregating all information retrieved in the previous steps. Moreover, after analyzing all the repositories, we performed an additional aggregation to have coarse-grained data that can provide meaningful insights. \red{We considered the repositories with at least 5 microservices.}

The final dataset~\footnote{The replication package along with the dataset is available at \url{https://github.com/deib-polimi/microservices-analysis}.} contains \red{$24$} repositories, \red{$24$} docker-compose files, \red{$172$} Dockerfiles, and \red{$408$} related dependency files.

\section{Results: Microservice Practice Explained} 
\label{sec:results}
Retrieved repositories helped us answer our research questions (Section~\ref{sec:rq}).

\begin{figure}
     \centering
     \begin{subfigure}[b]{0.34\columnwidth}
    	\includegraphics[width=\columnwidth]{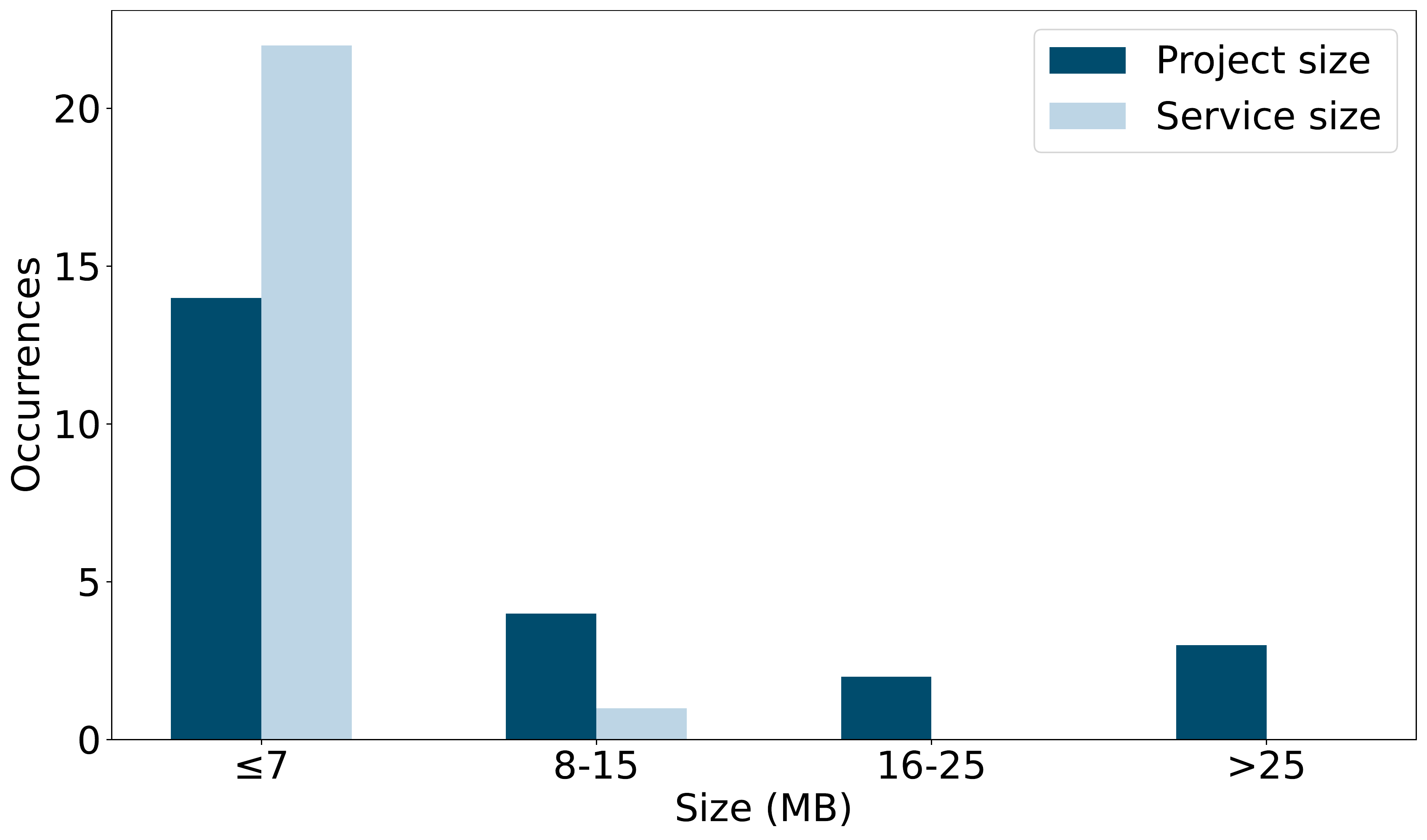}
       \caption{}
    	\label{fig:size}
     \end{subfigure}
     \begin{subfigure}[b]{0.34\columnwidth}
       	\centering
	\includegraphics[width=\columnwidth]{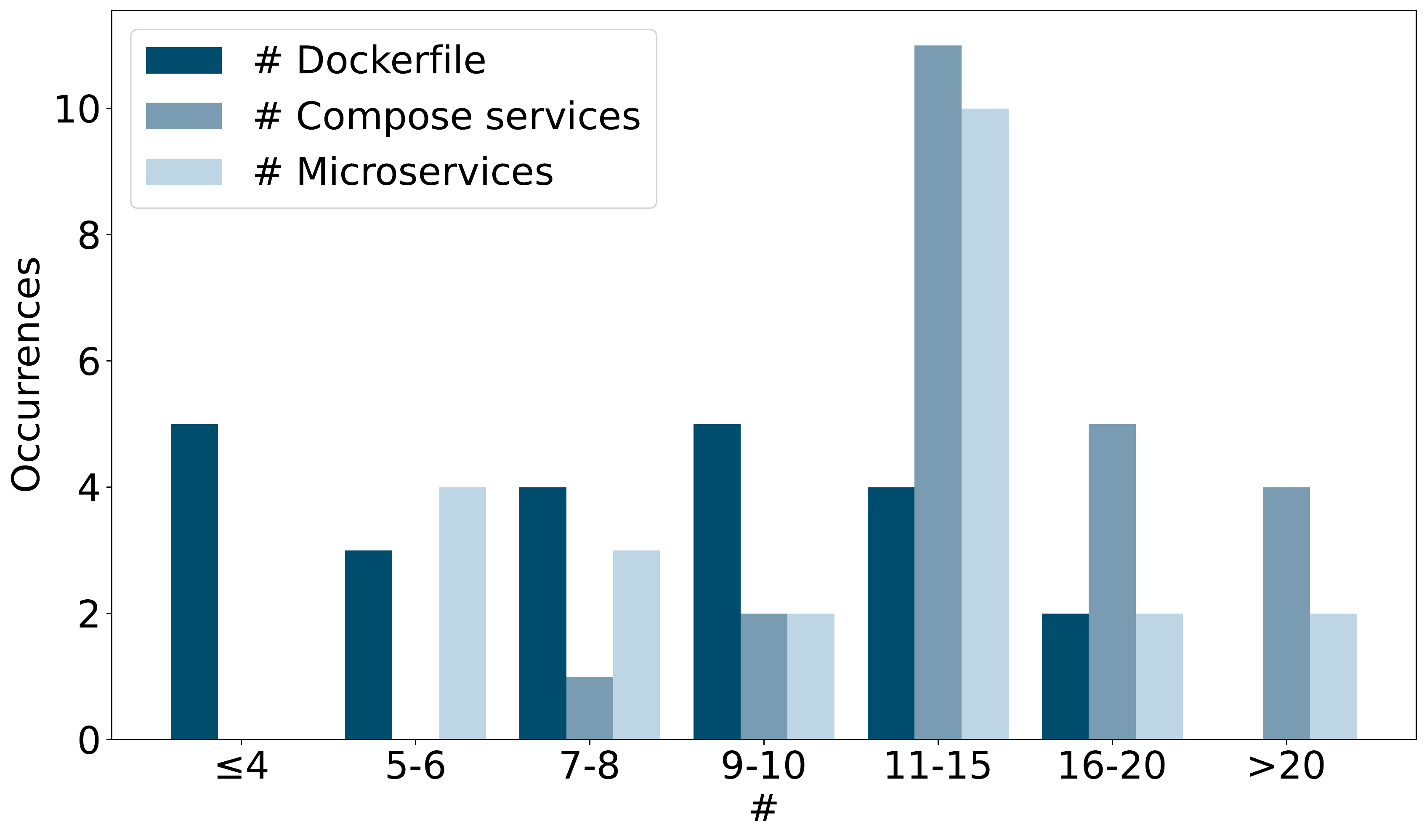}
		\caption{}
	\label{fig:serviceamount}
	\end{subfigure}
	  \begin{subfigure}[b]{0.30 \columnwidth}
	\includegraphics[width=\columnwidth]{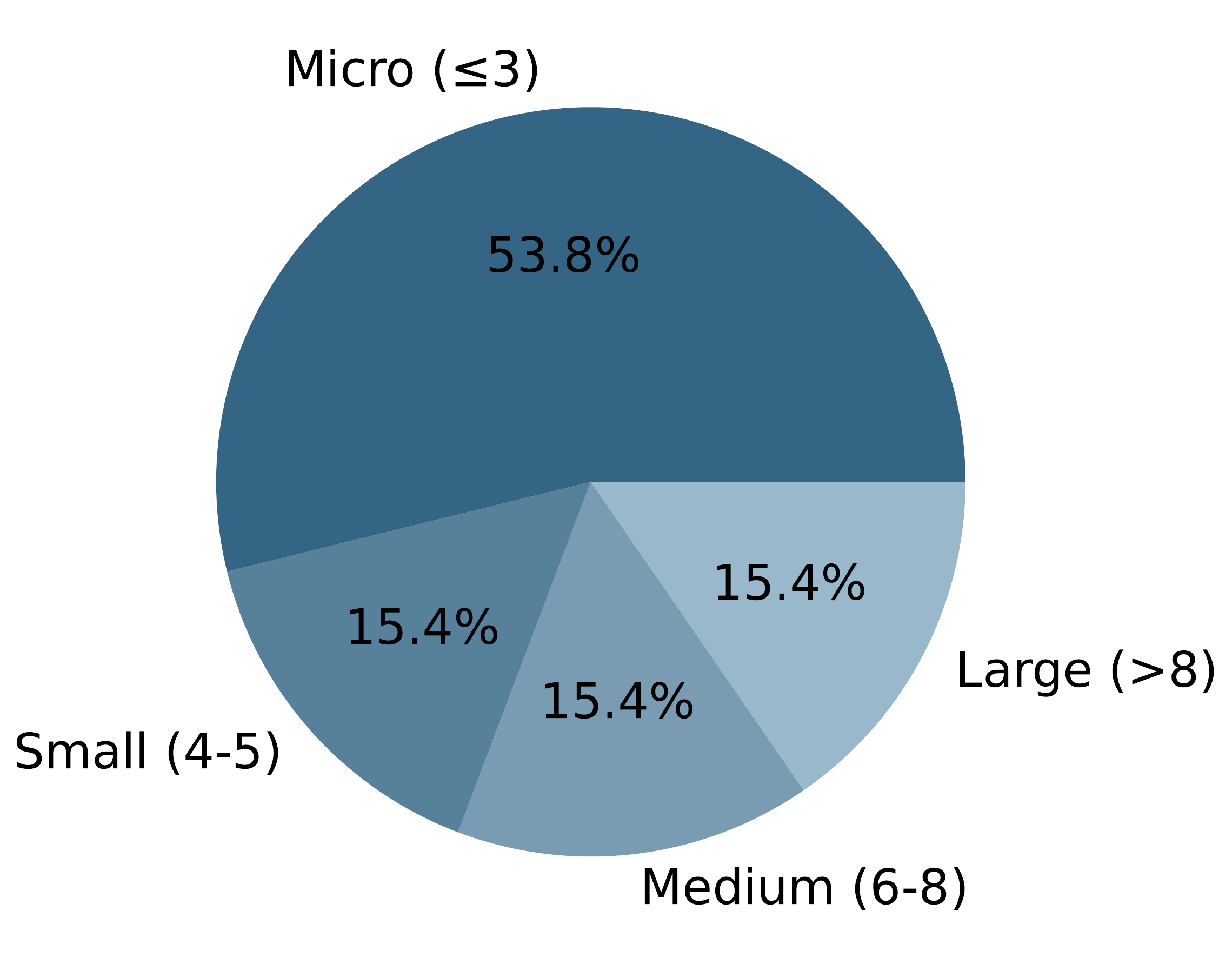}
	\caption{}
	\label{fig:teams}
     \end{subfigure}
        \caption{Project and microservice sizes (a), number of microservices per project (b) and team sizes (c).}
         \label{fig:sizetram}
\end{figure}

\noindent\textbf{\boldmath$RQ1$: Sizes.} To address the first research question, we first analyzed the size (in bytes) of each repository and of each microservice. Note we excluded intentionally databases, middleware infrastructures, and related components. Figure~\ref{fig:size} presents both the size of projects and the average size (per project) of identified microservices. We assume we have a microservice when a repository contains a folder with a Dockerfile and the corresponding source code, and when the Dockerfile is mentioned as a service image in the docker-compose file of the project. 

The figure shows that \red{all} the projects are under $50$MB, with an average of \red{$8.1$MB}, a standard deviation of \red{$10.7$MB} and the $95th$ percentile equals to \red{$30.77$MB}. On the other hand, the average size of microservices is around \red{$1.25$MB} with a standard deviation of \red{$1.73$MB}; $95\%$ of the services are under \red{$3.75$MB} and $80\%$ of the services are smaller than \red{$1.96$MB}.

The numbers say that there is no clear correlation between project size and number of contained microservices: bigger projects do not contain more services but simply bigger resources/artifacts. If we considered that all bytes refer to the source code, it would result in an estimated average of $31K$ lines of code. However, most of the repositories contain several resources, including images and documentation that could significantly increase the weight of a project. We can then infer that analyzed microservices are small enough and can be considered appropriate microservices.

Figure~\ref{fig:serviceamount} shows the number of Dockerfiles, compose services, and microservices. On average, we identified \red{$10.6$} microservices and \red{$14.2$} compose services per project. This indicates that on average each microservice corresponds to around ~$1.3$ compose service, which means that either some of the microservices are stateless, or that multiple companion elements (e.g., API gateways) are deployed along with the actual services, or that these microservices often share a database\footnote{Note that a stateful microservice comprises at least two compose services, one for the business logic and one for the database.}. This would be against the principle of \textit{independent data stores} mentioned in Section~\ref{sec:background}.

The number of Dockerfiles is close to the number of identified microservices, with an average of \red{$7.1$} services per project and a maximum of \red{$17$} (the maximum number of detected microservice is \red{$24$}). In the $75\%$ of the cases,  the projects contain at most \red{$13$} microservices, \red{$18$} compose services, and \red{$9$} Dockerfiles. Recalling the principle that microservices are built around functionalities, either most of the projects are simple/toy applications or available components are not full-fledged microservices and an additional reorganization is needed to fully adopt the architectural style. 

The aforementioned principles also say that microservices should be developed by small teams (a principle already highlighted in agile practices). Figure~\ref{fig:teams} shows the distribution of team sizes per microservice. For each microservice, we computed the average team size as the ratio between the total number of project committers and the number of detected microservices. More than \red{$68\%$} of the analyzed microservices have been modified by at most  $5$ developers with a maximum team size of \red{$11.7$} persons. This is coherent with the Amazon's rule: a dozen person team is the maximum allowed for managing a project.

\begin{mdframed}
	\noindent\textbf{Answer to RQ1.} $80\%$ of analyzed microservices can be considered \textit{small enough}. The average number of microservices per projects is \red{$10.6$} and $75\%$ of them contain at most \red{$13$} microservices.
\end{mdframed}
Even more interestingly, our study reports on the sizes with which microservice operations are realised. Specifically, our study shows the following:
\begin{mdframed}
\noindent\textbf{Key Finding 1.} Team sizes are as small as the microservices catered to, typically reflecting 3 or less microservice ``owners''.
\end{mdframed}
The aforementioned finding reflects a tendency of microservice designers and operators towards the so-called ``NoOps'' trend \cite{ZasadzinskiSBMC18}, which itself deserves further study.


\noindent\textbf{\boldmath$RQ2$: Polyglotism.}

\begin{figure}
     \centering
     \begin{subfigure}[b]{0.49\columnwidth}
    	\centering
	\includegraphics[width=\columnwidth]{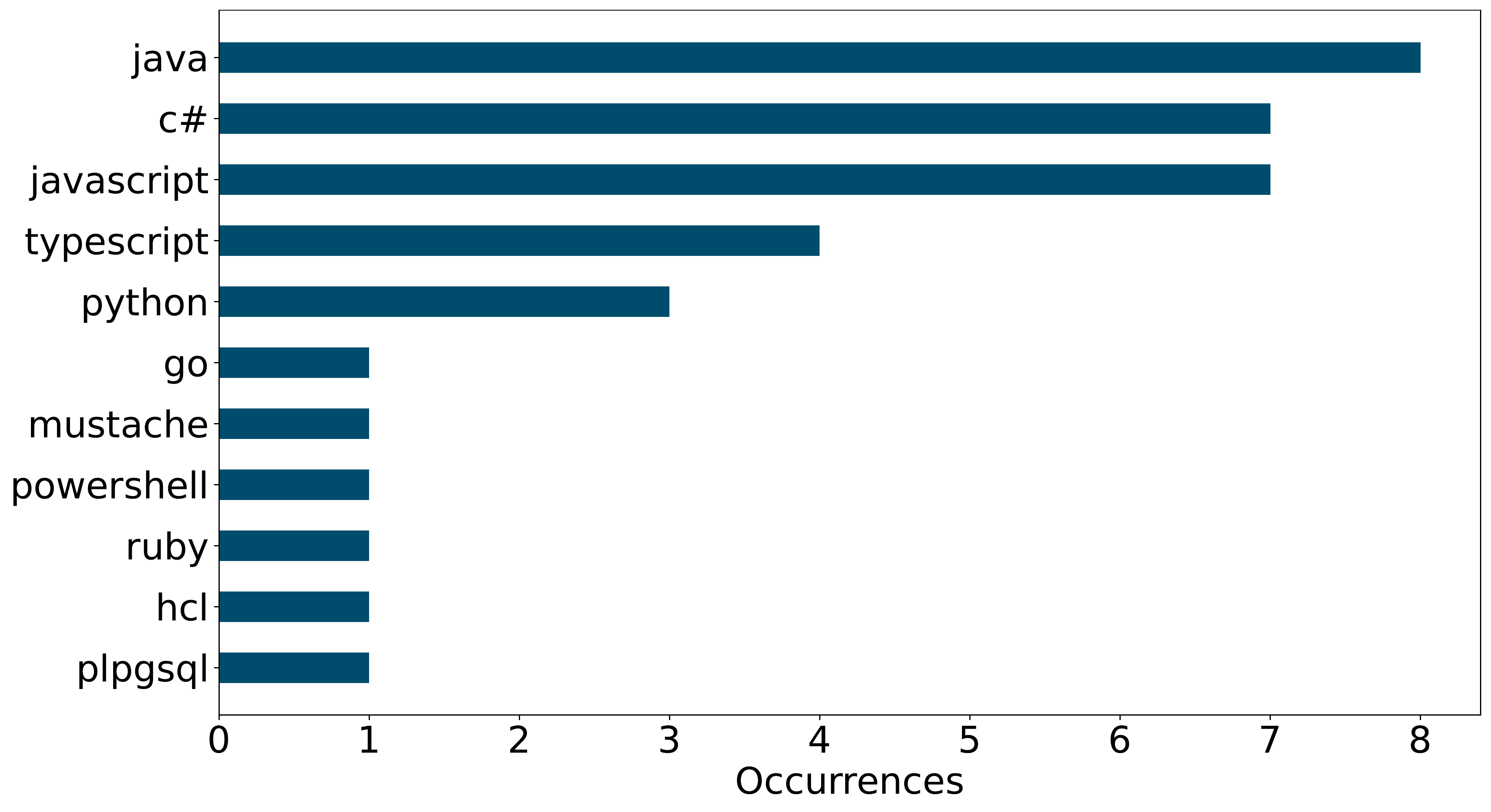}
	\caption{}
	\label{fig:langs}
     \end{subfigure}
     \begin{subfigure}[b]{0.49\columnwidth}
    	\centering
	\includegraphics[width=\columnwidth]{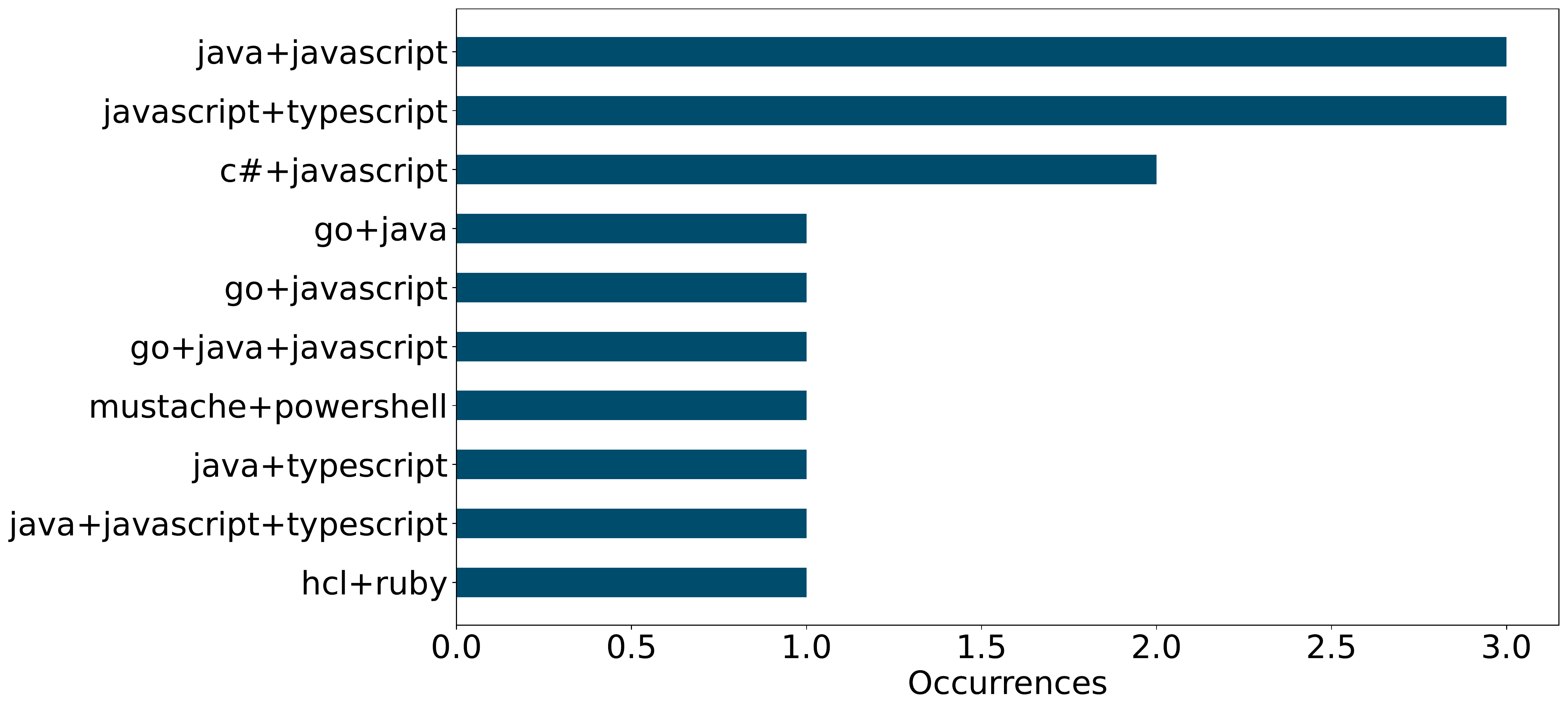}
	\caption{}
	\label{fig:langs-comb}
	\end{subfigure}
    \caption{Top-used programming languages (a), and top-used combinations of programming languages (b)}
         \label{fig:example}
\end{figure}

The analysis detected a total of \red{$9$} different languages.  Figure~\ref{fig:langs} shows the used programming languages. The top language is Java, which was utilized in \red{$8$} projects. Javascript resulted to be the second language (\red{$7$} occurrences), consistent with the measured popularity of \textit{node.js}.
C\# (\red{$7$} occurrences) appears to be more used than Python (\red{$3$} occurrences). 

To measure the adoption of multiple languages within a project (polyglotism), we compared the distribution of the number of microservices against the number of detected languages. On average each project uses $2$ languages with a maximum of \red{$4$}, but projects based on more than $1$ language (excluding markup ones) are about $40\%$ of the total. Since the average number of microservices is more than \red{$10$}, the ratio between microservices and languages per project is \red{$7$}. Note that a value equals to $1$ indicates that each microservice is written in a different language, while the greater the value is, the less polyglot the project is.

Some languages are also more used in combination than others. 
For example, Java was utilized alone in \red{$4$} projects (over the \red{$8$} occurrences), while Javascript only \red{$1$} times over \red{$7$} projects. This can be better visualized in Figure~\ref{fig:langs-comb} that shows the most widely used combinations of two or three languages. 
The most used combination is Java and Javascript with \red{$3$} occurrences. Javascript is also often used in combination with C\# and Typescript. The small number of combinations is caused by two factors: polyglotism is not common practice and Java is used in more than one third of the projects and it is rarely combined with other languages.

\begin{mdframed}
	\noindent\textbf{Answer to RQ2.} Microservices are built mainly in Java, Javascript \red{and C\#} with Polyglotism not emerging as a common practice. 
\end{mdframed}

At the same time, our evidence shows that several languages are combined more effectively with others. Finding and quantifying the effects of such patterns deserves a study of its own, e.g., to understand the effects and effect sizes caused by a specific combination of languages onto microservices execution times or scalability. In summary:

\begin{mdframed}
    \noindent \textbf{Key Finding 2.} There exist a specific set of languages and language types (e.g., Javascript), which emerge as more suited to be used in conjunction with other languages. 
\end{mdframed}

\noindent\textbf{RQ3 - Application servers and DBMSs.} 
\begin{table*}[t]
	\centering
	\scriptsize
\begin{tabular}{c|c|c|c|c|c|c|c}
\cline{2-7}
 & \textit{mongo} & \textit{redis} & \textit{mysql} & \textit{postgres} & \textit{postgresql} & \textit{solid} &  \\ \hline
\multicolumn{1}{|c|}{\textit{node}} & $6$ & $4$ & $3$ & $1$ & $1$ & $1$ & \multicolumn{1}{c|}{$16$} \\ \hline
\multicolumn{1}{|c|}{\textit{spring}} & $3$ & $3$ & $4$ & $1$ & $1$ & $1$ & \multicolumn{1}{c|}{$13$} \\ \hline
\multicolumn{1}{|c|}{\textit{express}} & $4$ & $3$ & $0$ & $1$ & $0$ & $0$ & \multicolumn{1}{c|}{$8$} \\ \hline
\multicolumn{1}{|c|}{\textit{flask}} & $2$ & $2$ & $0$ & $0$ & $0$ & $0$ & \multicolumn{1}{c|}{$4$} \\ \hline
\multicolumn{1}{|c|}{\textit{glassfish}} & $1$ & $1$ & $2$ & $0$ & $0$ & $0$ & \multicolumn{1}{c|}{$4$} \\ \hline
\multicolumn{1}{|c|}{\textit{jetty}} & $0$ & $1$ & $1$ & $0$ & $0$ & $0$ & \multicolumn{1}{c|}{$2$} \\ \hline
\multicolumn{1}{|c|}{\textit{tomcat}} & $0$ & $1$ & $1$ & $0$ & $0$ & $0$ & \multicolumn{1}{c|}{$2$} \\ \hline
\multicolumn{1}{|c|}{\textit{asp}} & $0$ & $0$ & $0$ & $0$ & $0$ & $0$ & \multicolumn{1}{c|}{$0$} \\ \hline
\multicolumn{1}{|c|}{\textit{jboss}} & $1$ & $1$ & $0$ & $0$ & $0$ & $0$ & \multicolumn{1}{c|}{$2$} \\ \hline
\multicolumn{1}{|c|}{\textit{monkey}} & $0$ & $0$ & $1$ & $0$ & $0$ & $0$ & \multicolumn{1}{c|}{$1$} \\ \hline
 & $17$ & $16$ & $12$ & $3$ & $2$ & $2$ &  \\ \cline{2-7}
\end{tabular}
	\caption{Auxiliary frameworks and DBMSs}
	\label{tab:stacks}
\end{table*}

The analysis detected \red{$10$} different application servers and \red{$6$} databases. On average, we identified that projects use \red{$1.4$} types of application servers per project and \red{$1.4$} DBMS type. Note that using a single DBMS type does not imply that all the microservices in the project use the same DMBS, but they can use independent instances of the same DBMS.

\red{Table~\ref{tab:stacks} shows the used frameworks as rows, and the used DBMS as columns.} Each cell indicates the number of occurrences of the combination of the two technologies (e.g., the combination \textit{flask}-\textit{redis} is used in \red{two} different projects). The most widely used  framework is \red{\textit{Node}}, again a \red{Javascript}-based framework, while \textit{mongo} is the most utilized DBMS. The most utilized pair is \textit{mongo}-\textit{node.js}.

The table also says that Spring is commonly used with relational databases such as MySQL and Postgres, while, \textit{node.js} is mostly used with \textit{mongo} and \textit{redis}. This witnesses the idea that a weakly typed language as Javascript fits naturally with schema-less data stores. Express\footnote{\url{https://expressjs.com}}, the third most commonly adopted application server is a library for \textit{node.js}, while \textit{flask} (\red{$4$} occurrences) is widely used in Python environments. 

\begin{mdframed}
	\noindent\textbf{Answer to RQ3.} \textit{node.js} and \textit{Spring}  are the most commonly used frameworks and they are usually paired with a NoSQL and a relational database, respectively.
\end{mdframed}

Since microservices are loosely coupled, they can use different technologies and platforms. However, our study shows that:

\begin{mdframed}
    \noindent \textbf{Key Finding 3.} 
    Multiple technology stacks are used only sporadically.
\end{mdframed}

\noindent\textbf{RQ4 - Topologies and Data Independency.}

We investigated the dependencies among microservices by creating a DAG for each docker-compose file. For each compose service, we created a node in the graph and for each \textit{depends\_on} entry we created a link with another node. The analysis allow us to understand how the components of a service are related and interdependent. We summarized the results in Table~\ref{tab:dep_graph}. The table shows the average number of nodes, edges of the graph, along with the average dependencies per services and the average and max longest path in the graph. The row called 'full' presents the results considering all the services in the docker-compose, while the 'micro' values are computed on the dependencies graphs with only applicaton-level logic (i.e. removing DBMS, servers, buses, etc. nodes). On average, the graphs are sparse and the longest paths are short that is, the services are not very interdependent. As expected, the graphs are acyclic, there are no circular dependencies in the graphs.
 
Finally, we studied the architecture topologies extracted from docker-compose files to understand whether database elements are shared among different microservices. As in the dependencies graph analysis, we created a node in the topology for each compose service, and a link with another node for each \textit{depends\_on} entry. If a compose service is a data store and has more than one incoming link in the topology, it is shared among different microservices. This case would clearly violate the principle of independent data stores. In total we measured only \red{$9$} projects (\red{37.5\%})  with at least one shared data store. Considering that sharing databases is an easy way to avoid inter-database communication, such a small number is surprising.

\begin{mdframed}
	\noindent\textbf{Answer to RQ4.} Microservices are in fact loosely coupled. The average number of dependencies among microservices is $0.62$. If we consider also connections with auxiliary components (e.g., databases or gateways) the average number of dependencies increases to $1.05$.
\end{mdframed}

Even more importantly, and with some extent of violation against the typical tenets associated with microservices design and computing, our data indicates that sharing DBs typically plays a role in the current microservice landscape. Specifically:

\begin{mdframed}
    \noindent \textbf{Key Finding 4.} \red{$37.5\%$} of the  projects under study in our sample features a database shared among different microservices.
\end{mdframed}

\noindent\textbf{RQ5 - Communication and management.}
\begin{figure}
     \centering
     \begin{subfigure}[b]{0.42\columnwidth}
    	\includegraphics[width=\columnwidth]{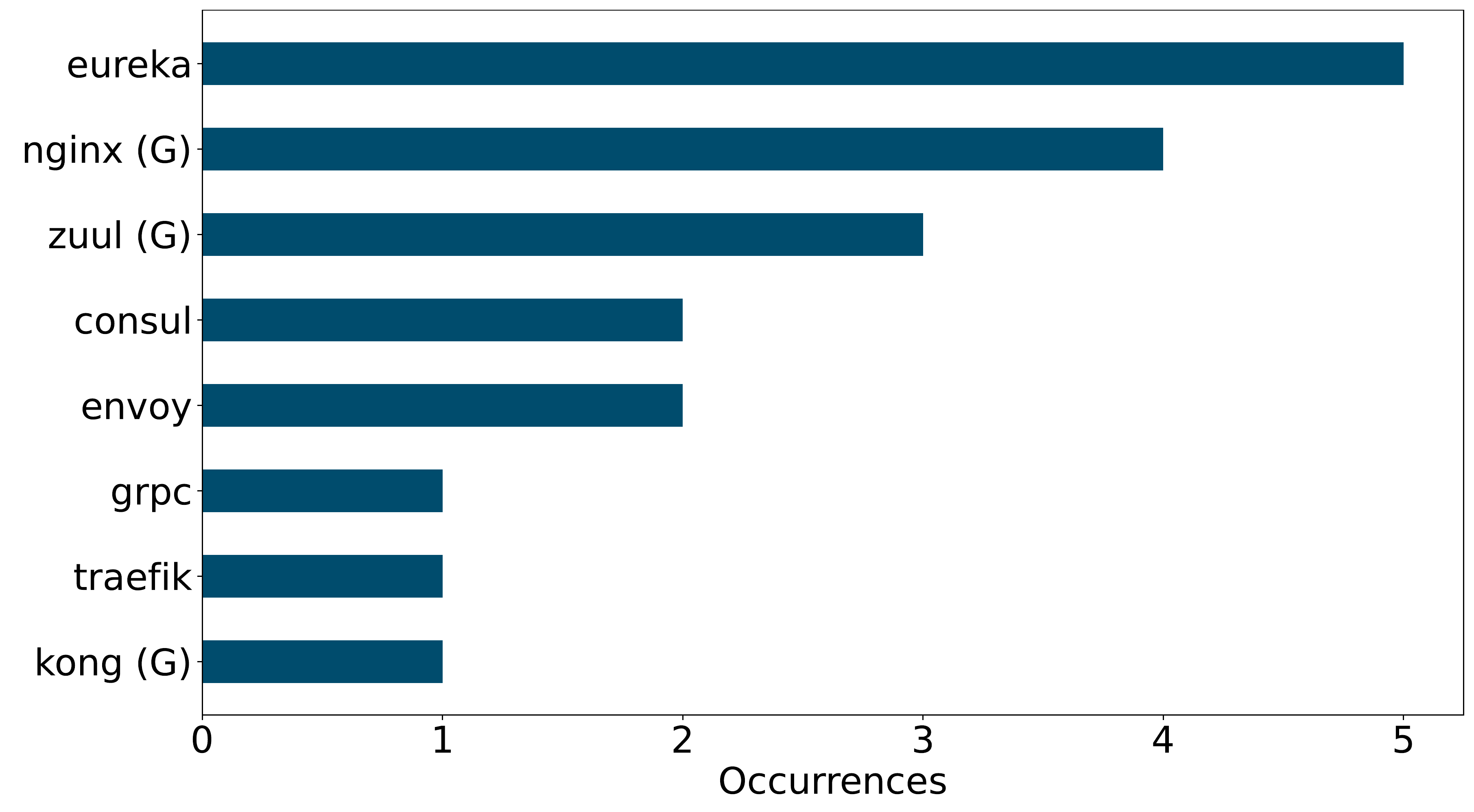}
	\caption{}
	\label{fig:gates}
     \end{subfigure}
     \begin{subfigure}[b]{0.42\columnwidth}
       	\centering
	\includegraphics[width=\columnwidth]{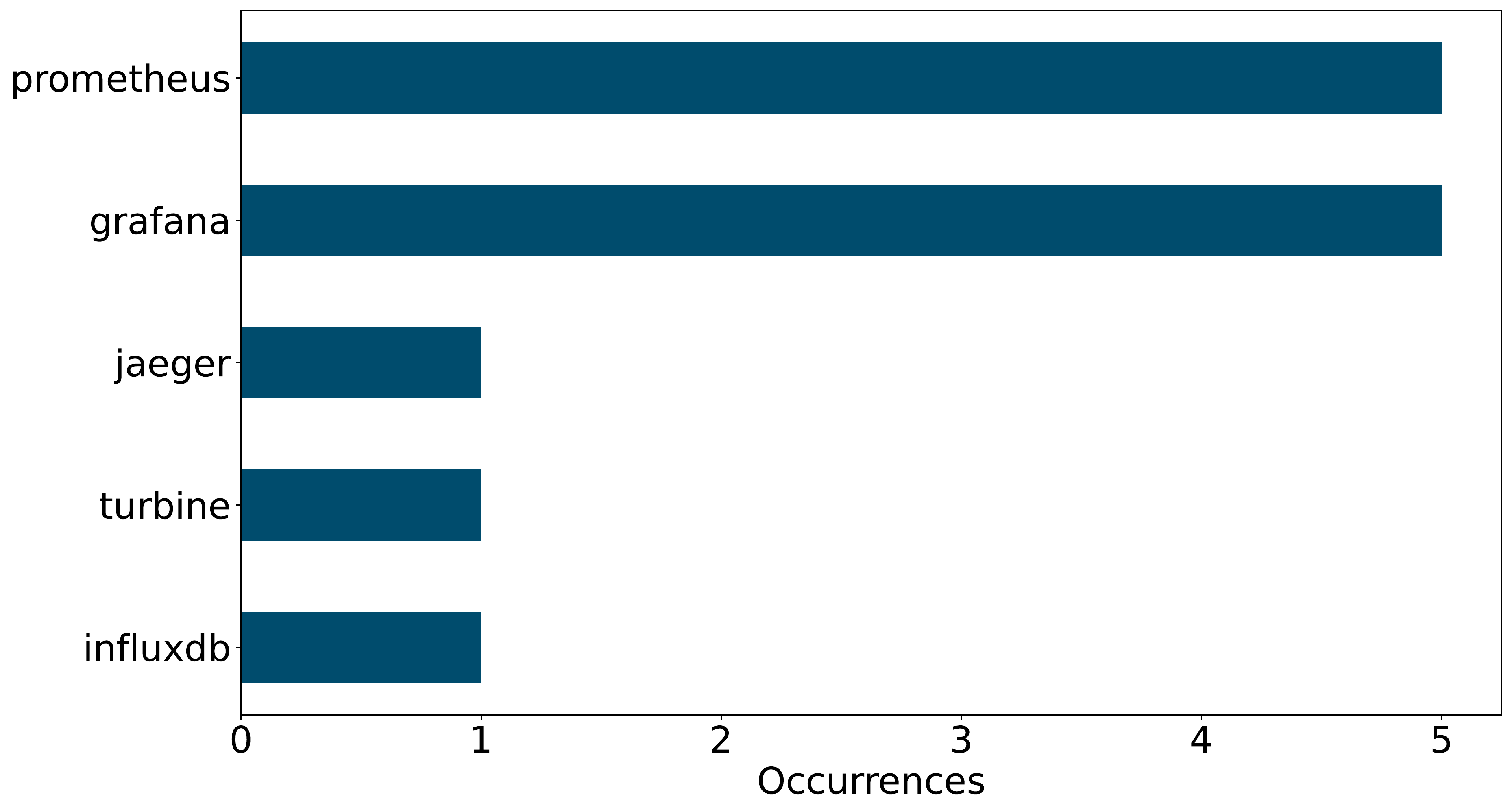}
	\caption{}
	\label{fig:monitors}
	\end{subfigure}
        \caption{Top-used API gateways (marked with G) and service registries (a), and top-used monitoring infrastructures (b).}
    \label{fig:aux}
\end{figure}

\begin{table*}[t]
	\scriptsize
	\centering
\begin{tabular}{c|c|c|c|c|ccc|}
\cline{2-8}
 & \textit{Nodes} & \textit{Edges} & \textit{Avg deps per service} & \textit{Num cyclic} & \multicolumn{3}{c|}{\textit{Longest path}} \\
 & avg & avg & avg &  & avg & min & max \\ \hline
\multicolumn{1}{|c|}{full} & 14.21 & 15.75 & 1.05 & 0 & 2.21 & 0 & 7 \\ \hline
\multicolumn{1}{|c|}{micro} & 9.96 & 6.79 & 0.62 & 0 & 1.62 & 0 & 6 \\ \hline
\end{tabular}
	\caption{Dependencies Graph Values}
	\label{tab:dep_graph}
\end{table*}

We measured the usage of lightweight message busses instead of bare HTTP/REST client-server interactions. Quite surprisingly, \red{more that half of the projects ($14$)} uses at least one bus infrastructure. \red{$5$} projects use \red{at least two} different buses, with an average of $0.8$ buses per project.  Most projects (\red{$11$}) use RabbitMQ, \red{or the Advanced Message Queuing Protocol (AMQP).}  These data seem to indicate that the simple HTTP client-server paradigm is often not enough or not sufficiently efficient either when microservices communicate asynchronously or  to implement service choreographies. 

As for microservice management, we measured how many projects employ an API gateway and/or a service registry. \red{$4$} projects use an API gateway and \red{$9$} projects a service registry; on average, we have \red{$0.4$} API gateways and \red{$0.4$} service registries per project. Only \red{$1$} repository contain \red{more than two API gateways, and one project} use more than one service registry. Figure~\ref{fig:gates} shows the most used technologies. Netflix Eureka is by far the most commonly used service registry with \red{$5$} occurrences, while \textit{consul}\footnote{\url{https://www.consul.io}} and \red{\textit{Envoy}\footnote{https://www.envoyproxy.io/}} both resulted in \red{$2$} matches. \textit{Netflix Zuul}\footnote{\url{}https://github.com/Netflix/zuul} is the top-used API gateway (\red{$3$} occurrences), followed by NGINX and Kong with \red{$1$} occurrence.  

These results suggest that it is often not enough to consider microservices as small REST-based services, but they embody a more complex structure. The architecture itself adds a layer of complexity that must be managed with dedicated solutions to ease the development and management of these systems. Note that cloud providers have started to offer API gateways and service registries. For example, Amazon Web Services provides these tools in its Elastic Container Service (ECS)\cite{awsdisco}. Thus, obtained data could even be a lower bound of how spread these practices are.

The analysis also highlights that several projects (\red{$7$}) embed an observability/monitoring component. Figure~\ref{fig:monitors} shows the top-used monitoring solutions. Prometheus\footnote{\url{https://prometheus.io}} and Grafana\footnote{\url{https://grafana.com}} (a companion tool for data visualization) are the first ones with \red{$5$} matches each, while \red{InfluxDB \footnote{\url{https://www.influxdata.com/}}, Jaeger\footnote{\url{https://www.jaegertracing.io/}}, and Netflix Turbine\footnote{https://github.com/Netflix/Turbine}} obtained $1$ preference. This is also witnessed by the presence of Prometheus and Grafana in the list of the most significantly used container images (see $RQ5$) 

\begin{mdframed}
	\noindent\textbf{Answer to RQ5.}  Message bus systems are adopted more than expected, with RabbitMQ is the top-used solution; \red{API gateways are used by only the $15\%$ of the repositories, while service registries are used by more than one third of analyzed projects.}  
\end{mdframed}

At the same time, our evidence reflects one more key finding, namely:

\begin{mdframed}
    \noindent \textbf{Key Finding 5.} Only in some \red{$30\%$} of the cases, a monitoring infrastructure is deployed within the system.
\end{mdframed}

This finding reflects previous evidence on the matter; specifically, Tamburri et al. \cite{TamburriMN20} discover that technical efforts dedicated into specific and effective monitoring infrastructure is scarce, stove-piped, and overall, in its infancy. More research into the effective use of monitoring patterns in combination with microservice architectural design is needed to quantify such effectiveness, especially in terms of benefits for microservice continuity \cite{NiemimaaJ13}.

\begin{figure}[t]
	\centering
	\includegraphics[width=0.45\columnwidth]{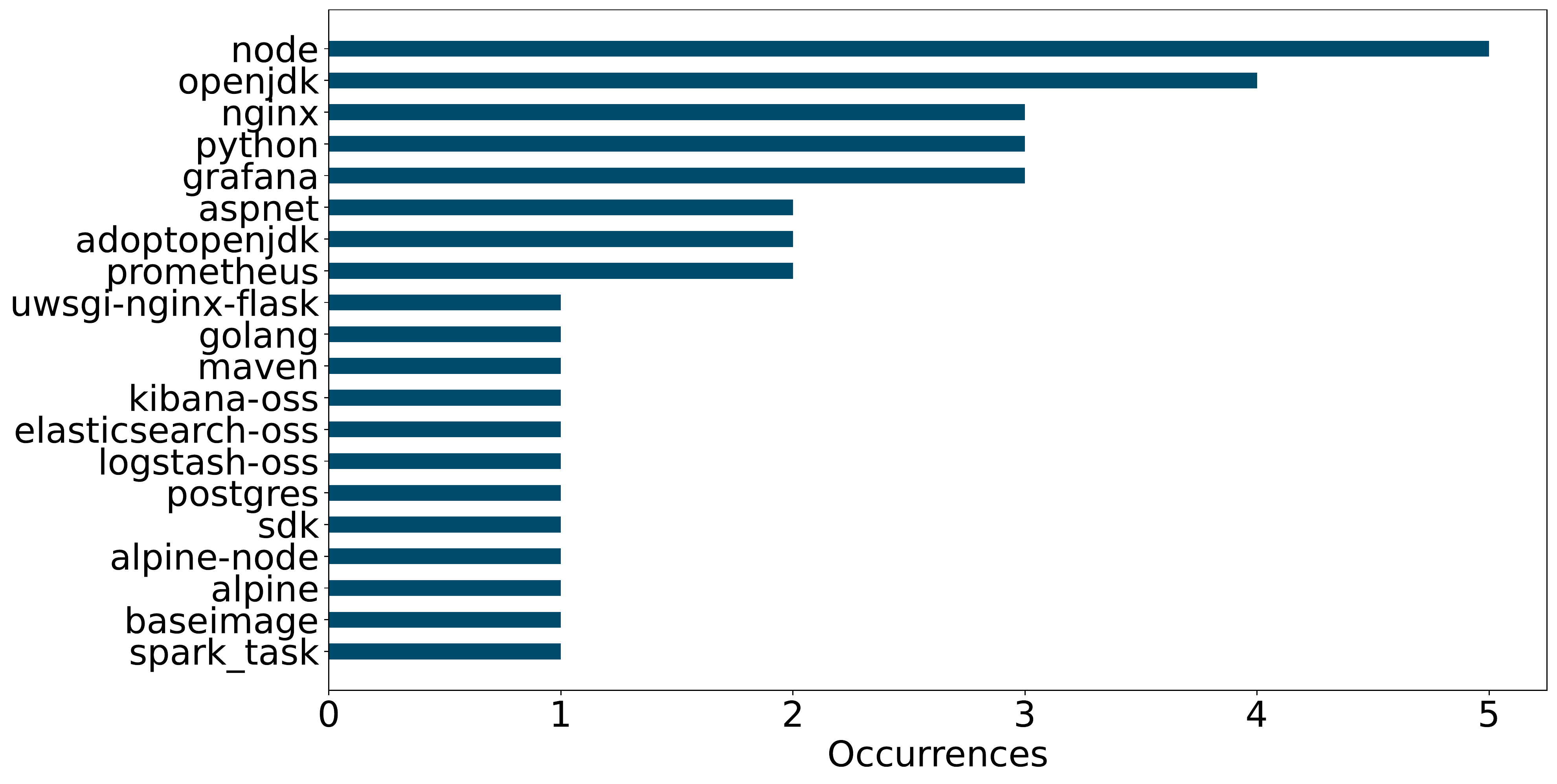}
	\caption{Top re-used container images.}
	\label{fig:images}
\end{figure}

\noindent\textbf{RQ6 - Reused container images.} Our analysis scanned a total of \red{$172$}  Docker configuration files and also retrieved used container images. Each Dockerfile defines one base image (instruction \textit{FROM}), while used, pre-built images are defined in docker-compose files through attribute \textit{image}. 
Figure~\ref{fig:images} shows the most widely re-used images. Note that each image is counted only once per project even if used in multiple microservices. These images could be both starting points for custom user code or components ready to be deployed. 

\red{\textit{node.js} is the most used image with $5$ occurrences}, showing the recent trend of using Javascript as server-side language. \red{Java \textit{openjdk} is in second position being used in $4$ projects.} In \red{third} place, \textit{NGINX} appears to be the standard de-facto API Gateway for microservices with $3$ occurrences. \red{Python is used $3$ projects.} \red{We identified an occurrence of \textit{alpine} but we did not find any of \textit{ubuntu} and \textit{debian}, images that only contain the operating system and that are commonly used to start from scratch and build a custom images with very limited pre-installed software.}

We also measured the most widely used combinations of two or three images, with \textit{NGINX+node} as first option (\red{$2$} matches). This is \red{not} quite surprising given the above result. \red{As for monitoring, \textit{openjdk} combined with \textit{Prometheus} is adopted in $2$ projects}. Another interesting relevant combination is \textit{Prometheus} and \textit{Graphana} (\red{$2$} matches). 

\begin{mdframed}
	\noindent\textbf{Answer to RQ6.} \red{Node} images are by far the most used ones, followed by the ones related to the \red{\textit{Java}} platform. On average, $2$ different images are used on each project and the most popular combination is \textit{node.js} and \textit{NGINX}.
\end{mdframed}

These small numbers are not fully significant from a statistical standpoint, but we must consider that multiple images are customized by users and the identification of proper matches becomes more difficult. However, these data seem to confirm what observed in the other research questions.

\begin{mdframed}
    \noindent \textbf{Key Finding 6.} Most container images used in microservice architectures are custom ones.
\end{mdframed}

\section{Discussion and Threats to Validity}
\label{sec:discussion}
The microservice architectural style is becoming the reference architecture for conceiving new complex systems or for modernizing existing ones.  However, as described by Lewis and Fowler~\cite{lewis2014microservices}, the knowledge about it is mostly gathered from experience and practitioners. While the lack of clear definitions and processes creates some challenges in grasping the actual notion of microservice, the research community can help define these practices (e.g., \cite{Meyer2014AgileT}), and shed some light on the actual state of the ecosystem. This study moves in the latter direction and provides some insights and observed trends.

Analyzed microservices seem to be relatively small and are often maintained by teams that do not include more than 12 developers. From a purely syntactic standpoint, this is consistent with the principle of being small and loosely coupled components. However, automated analyses can hardly decide whether considered microservices  are correctly built around functionalities or are just simple/toy examples. This additional step would require either manual code inspection, whose accuracy should then be verified, or specific, and probably narrow, assumptions on used language constructs. Team sizes are also coherent with the principles but, once again, the finding is limited to the contributors of GitHub repositories and there is no way to understand whether the team owns the different required skills. 

We can advocate that Docker configuration files are effective in capturing the technical trends and in understanding how the architecture is instantiated. We have discovered that Java, and the Spring framework, are leading the ecosystem. However, Java-written services seem to be less prone to be used with other technology stacks than those written in languages like Javascript and Go. No matter the programming language, polyglotism is more an ideal and nice-to-have characteristic rather than something that belongs to current practice. The reason behind this could be that when we are used to and skilled in some technologies, we are not eager to move to new ones since the cost of their adoption is high and discouraging. Note that even if all the microservices in a project are based on the same software stack, the main principles should not be violated. 

We also found more evidence than expected as for the use of asynchronous communication, based on lightweight message busses, API gateways, and service registries. In addition, the analysis revealed that in most of the cases each microservice has its own data store, and thus the isolation from and independence of the others is preserved.  This also justifies the need for dedicated message busses to better serve the distributed transactions among the different data stores~\cite{vstefanko2019saga}.

\subsection{Threats to validity}
\label{sec:threats}
The threats~\cite{wohlin2006empirical} that may invalidate our results are:

\textbf{Internal validity}. We initially retrieved data from GitHub in a completely automated way. While selecting highly rated and active repositories removed most of the ``noise'', blindly assuming we had proper repositories about microservices could have threaten the internal validity of our study. Before manually filtering the dataset, we analyzed the results and obtained some unrealistic findings. For example, we had that the most widely used combination of languages was $C$ and Go even if the $C$ language is almost unused for this kind of applications. To mitigate this issue, we carried out a manual filtering step, where we discarded more than $50\%$ of the initial repositories (as described in Section~\ref{sec:methods}) since they were not instances of microservice architectures. The manual filtering allowed us to discover that $C$ and Go are used for building tools for microservices and not for implementing microservices.

\textbf{External validity}. Obtained results may not be easily generalized. The main problem is the assumption behind this work that GitHub is a significant source of repositories related to microservice architectures. While we have already described how we mitigated the possibility of incurring in false positives (repositories not related to microservice architectures), herein the issue lies in understanding that selected projects are good representatives of the industrial practice. We selected some proper applications, but most of the applications were exemplar ones. In general, since microservices are related to commercial products rather than tools, it is uncommon that these applications are open sourced (e.g., Netflix shares many tools related to microservices but, obviously, not its own applications). 
In the future, we plan to extend this work and further validate the results through interviews with practitioners and other sources of truth.

\textbf{Conclusion validity}. The main threats to our ability of drawing conclusions from results is the size of the final examined dataset. Starting from about $17,000$ matches, we discard more than 99\% of the results by using the described automatic and manual filters. However, we were still able to examine more than $150$ repositories and almost $800$ Docker configuration files. This makes the results statistically significant. 

\textbf{Construct validity}: While some characteristics of microservice architectures are easily measurable (e.g., polyglotism, shared data stores, usage of message buses), others are more blurry. For example, the size of a microservice cannot easily be measured since there is no straightforward operationalization of the concept.  In this case, we think the problem lies in the lack of a quantitative definition rather than in the experiments we carried out.

\section{Related Work} 
\label{sec:related}
Microservices have been the target of many studies in the last years. Taibi et al.~\cite{taibi2018definition} identify and discuss the most common bad practice when developing  and managing microservices. They start from a set of well-known bad smells and conduct a interview-based survey to classify and validate them. They discover that the most frequent issue is to have an improper split of microservices, that is, the boundaries among the functionality addressed by each service overlap. Consequently, they also report that the principle of independent data stores is also often violated ($41\%$ of the cases). According to our study, this issue is only present in $13\%$ of analyzed repositories. This difference could be caused by the fact that some of the repositories we analyzed are exemplar applications and thus they should follow all the known principles. 

Kalske et al. \cite{kalske2017challenges}, Di Francesco et al.~\cite{di2018migrating}, and Balalaie et al. \cite{balalaie2018microservices} discuss the challenges of converting existing applications (i.e., a monolith) into microservice  architectures. The technical challenges include the ability of understanding the root cause of failures, performance issues, and the management of the communication among services. While these problems can be faced also in other architectural styles, the distributed nature of microservices makes them more complex to address. Our study witnesses a significant usage of tools that simplify these issues: message busses often complement simple HTTP client/server communications, API gateways and service registries seem well-established components that ease architecture management, and monitoring systems are often employed. \cite{balalaie2018microservices} also highlights how containerization is key for migrating existing systems and managing these architectures. 
\cite{kalske2017challenges} also highlights that organizational challenges are more complex and subtle to tackle  and automation (continuous integration and delivery) must be embraced.

Zhang et al.\cite{8703917} carry out an empirical study on microservices that compares the principles defined by Lewis and Fowler~\cite{8703917} with the real state of practice. To do that they interviewed $13$ practitioners of different companies. The authors claim that with the lack of clear guidelines, some of the principles could cause significant problems to a company. For example, if microservices are not designed to be really independent, the updates of one service can disrupt the others or can increase the difficulty of testing and debugging. Moreover, if the organization does not catch up in adapting to the architecture,  the communication within the organization and among the different teams could rapidly increase. According to their interviews the most used language for producing microservices is Java as we measured in our study. However, they also claim that $3$ practitioners out of $13$ pointed out that the company suffered from excessive heterogeneity in the software stacks,  and that, in general, multiple languages were used. Similarly, we infer that polyglotism is not common practice. However, it is not clear from their study if the different languages were used only for developing microservices or for other tasks.

While the aforementioned studies used interviews for conducting the empirical studies, Baresi et al.~\cite{baresi2020microservices} used GitHub repositories as we did. They only focused on some simple elements of the repositories in general without focusing on Docker configuration files, thus their analyzed dataset is larger than ours. They also filtered more than 98\%  of the data resulting in a dataset of $651$  repositories. While they also notice some unrealistic results such as the presence of tools related to microservices and not microservices themselves, they did not rely on an additional manual filtering as we did.
They analyzed the data by performing a textual search in the  README files of the repositories and carried out a topic-based search, in order to understand some trends related to microservices. Obtained results include some technical insights: they show, as we do, that the most used programming language is Java and interestingly they discover that the second most common topic in microservice-related projects is Docker, which supports our work. However, they did not investigate the actual code, and their analysis is not aware of the actual number of microservices in each repository, of how they are connected, and of all the metadata hiden in the configuration files.

Finally, Cito et al. \cite{cito2017empirical} propose an empirical study about the Docker ecosystem in Github. They provide insights on the popularity of used programming languages and on the most used Docker images and instructions. They focused on Dockerfiles and did not consider docker-compose files. As we did, they analyzed the dependencies defined in the Dockerfiles through command \textit{RUN} and they propose to introduce a more abstract command to explicitly define the dependencies: this would clearly have simplified our work. They also evaluated the quality of Dockerfiles by measuring how many of them lead to a successful build of the image and how many best practices are followed by developers. A future extension of our work could also consider the quality of analyzed Docker configuration files to understand whether our dataset diverges from the findings of these authors.

\section{Conclusions} 
\label{sec:conclusions}
More and more modern, complex software systems are based on microservice architectures even if their characterizing elements are not always well defined. This paper tries to fill this gap and presents the results on an empirical study that analyzes the Docker configuration files found in microservice-related GitHub repositories. On average, available projects are on the small side. Some results witness the well-known principles and common practice; others do not. Additional (industrial) projects are needed for a more complete assessment. Retrieving additional relevant projects to complement this analysis is our next step.

\section{Acknowledgments}
This work has been partially supported by the SISMA national research project (MIUR, PRIN 2017, Contract $201752ENYB$).
We would like to thank Dr. Nicholas Rasi for his contribution. 

\bibliography{main}

\end{document}